\documentclass[conference]{IEEEtran}
\usepackage{amsmath,graphicx}
\usepackage{cite}
\usepackage{graphicx}
\usepackage{epsfig}
\usepackage{psfrag}
\usepackage{subfigure}
\usepackage{url}
\usepackage{stfloats}
\usepackage{amsmath}
\usepackage{color}
\usepackage{amssymb}
\usepackage{algorithm}
\usepackage{algorithmic}
\usepackage{multirow}
\usepackage{array}

\title{Training-Free Non-Intrusive Load Monitoring of Electric Vehicle Charging with Low Sampling Rate}
\author{\IEEEauthorblockN{Zhilin Zhang$^*$, Jae Hyun Son, Ying Li, Mark Trayer, Zhouyue Pi}
\IEEEauthorblockA{Samsung Research America -- Dallas\\
1301 E. Lookout Drive, Richardson, TX 75082, USA\\
$^*$Email: zhilinzhang@ieee.org}
\and
\IEEEauthorblockN{Dong Yoon Hwang, Joong Ki Moon}
\IEEEauthorblockA{Smart Home Solution Lab\\
Samsung Electronics Inc.\\
Suwon, Kyeong-gi-do, Korea} }

\begin{document}
%
\maketitle
\begin{abstract}
Non-intrusive load monitoring (NILM) is an important topic in smart-grid and smart-home. Many energy disaggregation algorithms have been proposed to detect  various individual appliances from one aggregated signal observation. However, few works studied the energy disaggregation of plug-in electric vehicle (EV) charging in the residential environment since EVs charging at home has emerged only recently. Recent studies showed that EV charging has a large impact on smart-grid especially in summer. Therefore, EV charging monitoring has become a more important and urgent missing piece in energy disaggregation. In this paper, we present a novel method to disaggregate EV charging signals from aggregated real power signals. The proposed method can effectively mitigate interference coming from air-conditioner (AC), enabling accurate EV charging detection and energy estimation under the presence of AC power signals. Besides, the proposed algorithm requires no training, demands a light computational load, delivers high estimation accuracy, and works well for data recorded at the low sampling rate 1/60 Hz.  When the algorithm is tested on real-world data recorded from 11 houses over about a whole year (total 125 months worth of data), the averaged error in estimating energy consumption of EV charging is 15.7 kwh/month (while the true averaged energy consumption of EV charging is 208.5 kwh/month), and the averaged normalized mean square error in disaggregating EV charging load signals is 0.19.
\end{abstract}

\section{Introduction}
\label{sec:intro}

Non-intrusive load monitoring (NILM) or non-intrusive appliance load monitoring (NIALM) is an important solution to realize smart-grid and smart-home energy management benefits. It aims to estimate operation status and energy consumption of individual electronic appliances by monitoring aggregated current/voltage/power signals in the main circuit panel of a house or a building \cite{hart1992nonintrusive,zeifman2011nonintrusive,zoha2012non}.

Electric vehicle (EV) charging is becoming an important load element for smart grid analysis \cite{clement2010impact,PecanReportEV,luo2013} although home charging EVs recently entered the market. Due to the growing number of the EV customers, a utility might start to experience non-marginal impacts on parts of its distribution system.  Particularly, the gravity of this impact will depend on at what time, for how long, with what utility rate, and in what season these EVs are being charged \cite{PecanReportEVData}. Therefore, it is necessary to solicit the importance and urgency of monitoring EV charging load via energy disaggregation.

Another usage of monitoring EV charging load is to provide house owners the monthly energy consumption of EV charging. This monthly feedback information can help house owners in bill-management and travel-management in the same way as monthly gas bill and conventional monthly electricity bill \cite{hayes1981reduction}.

There are many algorithms available for the energy disaggregation of various residential appliances \cite{hart1992nonintrusive,NN2006,zeifman2011nonintrusive,zoha2012non,parson2012non,kolter2012approximate,kolter2010energy,johnson2012bayesian,Du2013}, such as hidden Markov model(HMM) algorithms \cite{parson2012non,kolter2012approximate,johnson2012bayesian}. However, these algorithms were not specifically designed for EV charging, and they require extensive training and a large computational load. Therefore, for practical implementation where simultaneously monitoring tens of thousands houses is required, those algorithms may not be an attractive solution.


In this paper, a novel algorithm for energy disaggregation of EV charging is presented. It has several desired advantages. (1) It can mitigate the interference coming from air-conditioner (AC) power signals. Thus, it could be very helpful for smart grid load analysis and management during  peak load time in summer. (2) It does not require training, which is an highly attractive feature toward practical implementation. (3) It demands a light computational load, thus suitable for monitoring tens of thousands residual houses in large scale. (4) It works well for data sampled at 1/60 Hz, which aligns with the data provision capability of many smart-meters. Experiments based on real-world power data showed that it exhibits far better performance than state-of-the-art algorithms.


\begin{figure}[t]
\begin{minipage}[b]{.48\linewidth}
  \centering
  \centerline{\epsfig{figure=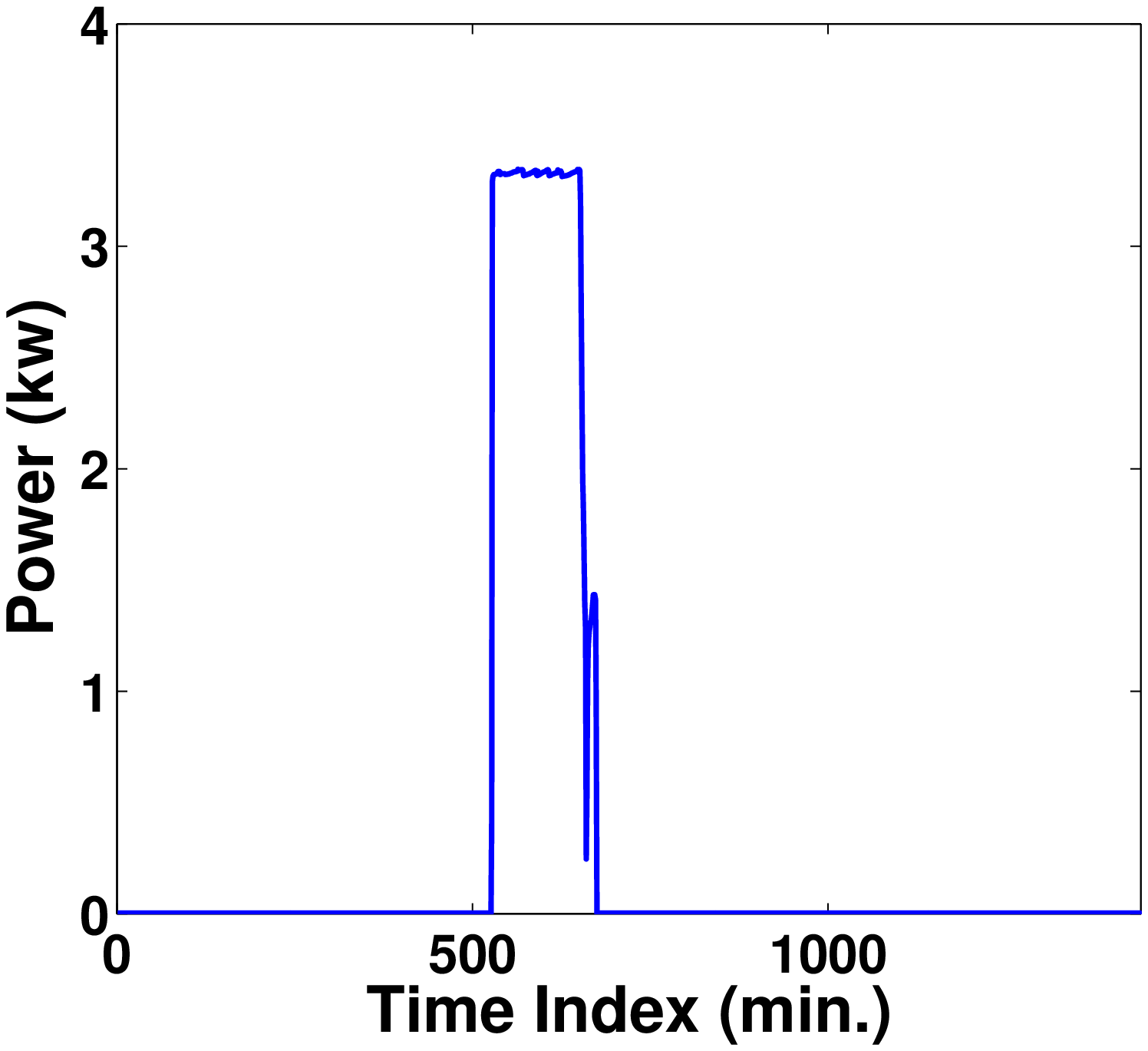,width=4.2cm,height=3cm}}
  \centerline{\footnotesize{(a) EV Power Signal}}
\end{minipage}
\hfill
\begin{minipage}[b]{0.48\linewidth}
  \centering
  \centerline{\epsfig{figure=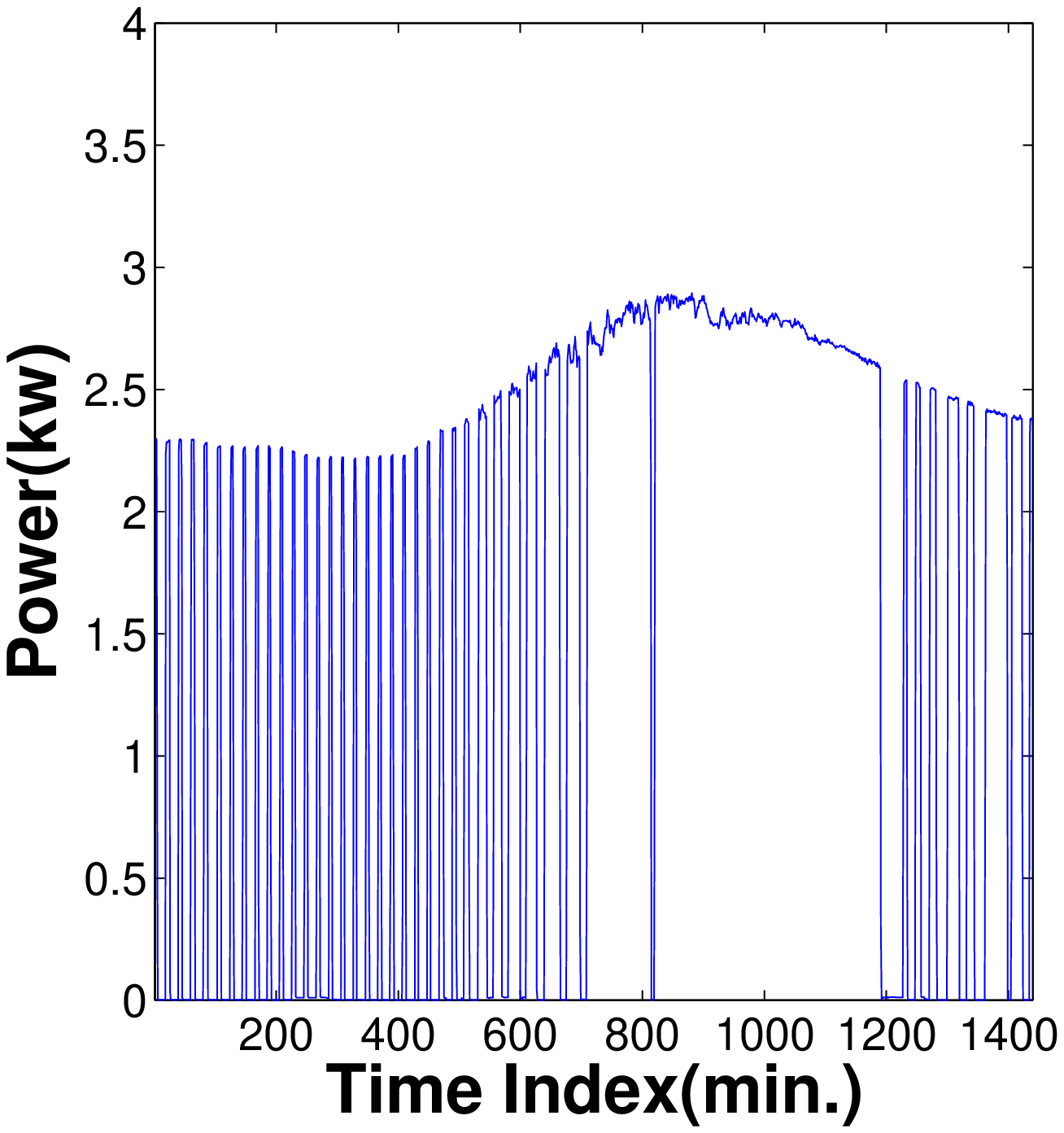,width=4.2cm,height=3cm}}
  \centerline{\footnotesize{(b) AC Power Signal}}
\end{minipage}
\caption{(a) An EV power signal. (b) An AC power signal exhibiting two kinds of waveform patterns, i.e. spike trains and lumps.}
\label{fig:ACnEV}
\end{figure}

\section{Challenges}
\label{sec:AC}

One big challenge of disaggregating an EV charging load from aggregated power signals is mitigating  interference  from AC. As shown in Fig.\ref{fig:ACnEV}(a), an EV charging load signal can be characterized as a square wave of a high amplitude (higher than 3 kW) and a long duration (longer than 30 minutes but generally shorter than 200 minutes) \cite{PecanReportEV}. AC power signals usually exhibit two kinds of waveform patterns. One pattern resembles a \emph{spike train} with very short durations (e.g. the train of waves from the 1st to 700-th minutes in Fig.\ref{fig:ACnEV}(b)). Another waveform pattern resembles a rectangular waveform of a high and slowly fluctuating amplitude and a long duration (e.g. the two lumps from the 700-th to 1200-th minutes in Fig.\ref{fig:ACnEV}(b)). This waveform pattern can seriously affect disaggregation performance of EV charging load signals due to the difficulty of distinguishing the AC waveform pattern from EV charging load signals, especially in the presence of other appliances' power signals and highly fluctuating residual noise. For notational convenience, this kind of AC waveforms will be called as  \emph{AC lumps}.

Another challenge lies with the aggregated data themselves of being real power signals sampled at 1/60 Hz. At this sampling rate, many useful appliance signatures \cite{laughman2003power,liang2010load,xia2012adaptive} such as transient characteristics available from high sampling rates no longer exist, which limits pattern recognition tools to render accurate  disaggregation results.

The third challenge is the lack of ground-truth of EV charging load signals for each house and  its large variation across different houses. To obtain the ground-truth of EV charging load signals in a given house, it requires to install sub-meter sensors to record these signals. However, it is unpractical to install such sub-meter sensors in every house. Thus, when disaggregating EV charging load from aggregated power signals in a given house, there is no training set (i.e., a collection of ground-truth of EV charging load signals in the house) available to train an algorithm. On the other side, EV charging load signals have large variation across different houses. For example,  EV charging load signals could have different amplitudes (although always higher than 3 kW), different width (i.e., charging duration), and different appearance time from house to house. As a result, an algorithm working well for a given house may perform poorly for another house.

In summary, a practical algorithm should work well for various houses and every season (especially the summer), and should not require training sets. But due to the above issues, to achieve high disaggregation accuracy of EV charging load is truly challenging.

\section{Proposed Algorithm}
\label{sec:EV}

\begin{figure}[t]
\centering
\includegraphics[width=8.5cm,height=7.0cm]{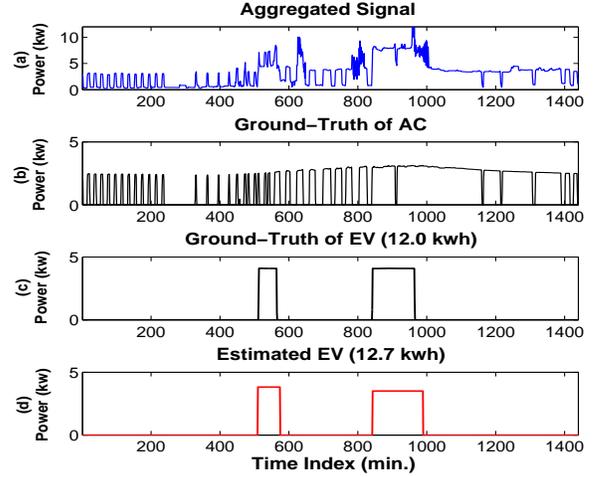}
\caption{\footnotesize{Energy disaggregation by the proposed algorithm. (a) An aggregated signal of one day. (b) The ground-truth of AC. (c) The ground-truth of EV (energy consumption is 12.0 kwh). (d) The estimated EV power signal (estimated energy consumption is 12.7 kwh). The energy estimation error (defined in Section \ref{sec:experiments}) is $5.8\%$, and the MSE is $0.178$.}}
\label{fig:EVest}
\end{figure}

\begin{figure}[t]
\centering
\includegraphics[width=8.5cm,height=8.5cm]{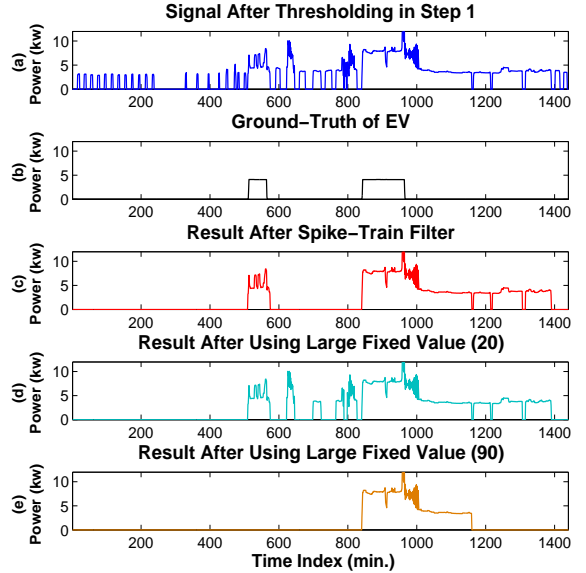}
\caption{\footnotesize{Results after using the spike-train filter and the fix-value thresholding method. (a) The aggregated signal after thresholding in Step 1. (b) The ground-truth of EV power signal, which has two waves. (c) The result after using our spike-train filter. (d) The result after thresholding using a small fixed value (20). There were many short AC spikes unremoved. (e) The result after thresholding using a large fixed value (90). Note that the first EV wave (from the 510-th to 550-th minute) was also removed.}}
\label{fig:bumptrainfilter}
\end{figure}

This section will describe the proposed algorithm  in detail. A one-day aggregated power signal (taken from the Pecan Street Database \cite{PecanStreetDatabase}) is chosen for illustration purpose (see Fig.\ref{fig:EVest}(a)).

\subsection{Step 1: Thresholding the Aggregated Signal}

For a given aggregated signal $x(t)$, first, a threshold $T_{\mathrm{low}}$  is applied to obtain a rough estimate of the EV charging load signal:
\begin{eqnarray}
\underline{x}(t) = \left\{
\begin{array}{ll}
x(t) & \quad x(t)\geq T_{\mathrm{low}}\\
0 & \quad x(t) < T_{\mathrm{low}}\\
\end{array} \right.
\end{eqnarray}
where $T_{\mathrm{low}} \triangleq \max\big\{2.5,  \frac{1}{2|x(k)>2|}\sum_{k: \; x(k)>2} x(k)\big\}$,
and $|x(k)>2|$ counts the number of sampling points whose amplitude is larger than 2 kW. After the initial thresholding (see Fig.\ref{fig:bumptrainfilter}(a)), the segments information of $\underline{x}(t)$ can be obtained such as the locations of a start-point and an end-point of each segment.

\subsection{Step 2: Filtering the Spike-Train}

Many AC spikes are present in the thresholding result (Fig.\ref{fig:bumptrainfilter}(a)), which need to be removed. One may consider setting a threshold to remove all spikes whose duration is shorter than the threshold. However, it is not easy to find a suitable duration threshold to remove all these spikes due to the varying nature of an AC spike duration (see Fig.\ref{fig:EVest}(b)).

Note that the duration of AC spikes gradually increases from morning to later afternoon and gradually decreases from later afternoon to midnight. Based on this observation, the following filter is designed to remove these spikes.

It first finds segments with duration shorter than $T_{\mathrm{seed}}=20$ (minutes), which are called `seeds' and labeled as `\emph{spikes to remove}'. Then, from each `seed', the filter searches the nearest segment \emph{forwardly}, checking whether the segment's duration is shorter than $D\triangleq(1+\eta) D_{\mathrm{cur}}$ and whether the gap between the `seed' and the nearest segment is no more than $3D_{\mathrm{cur}}$, where $D_{\mathrm{cur}}$ is the duration of the current `seed' and $\eta$ is a duration extension parameter ($\eta=1.2$ in our algorithm).  If this search condition meets, this nearest segment will be labeled as `\emph{spikes to remove}' and will be set as a new `seed'. Now, using this new `seed', the filter repeats the same \emph{forward} segment searching to the nearest segment, checking its condition in the same prescribed manner. If the search condition is not met, then jump to another `seed' and check its nearest segment forwardly as before. Similarly, the filter searches \emph{backwardly} as well. In the end, after completing the whole search range, all segments labeled as `\emph{spikes to remove}' are removed from $\underline{x}(t)$.

To prevent from removing a segment with a very large duration, one can adjust a threshold $T_{\mathrm{spike}}$ such that all removed segments have duration no more than $T_{\mathrm{spike}}$. For the proposed algorithm, it is set as $T_{\mathrm{spike}}=90 (\mathrm{minutes})$.

Note that the filter does not remove \emph{all} segments which have duration no more than $T_{\mathrm{spike}}$. It removes a segment only if its duration does not increase sharply compared to its surrounding segments' duration. If a segment with a long duration is surrounded by very short segments, even if this long segment has duration shorter than $T_{\mathrm{spike}}$, it will not be removed. The reason is that this segment could potentially indicate a waveform of EV, dryer, or oven.  So, it requires  further examination.

Fig.\ref{fig:bumptrainfilter} shows one example that the proposed spike-train filter removed all AC spikes. While using a fixed threshold value to remove these AC spikes, several single spikes were unremoved due to a small threshold (Fig.\ref{fig:bumptrainfilter}(d)), or a part of an EV power signal was removed mistakenly due to a large threshold (Fig.\ref{fig:bumptrainfilter}(e)).

%
%
%
%

\subsection{Step 3: Removing Residual Noise}

Residual noise refers to the mixture of errors from fluctuation of power signals, loss in power lines, and power signals of appliances with a low amplitude. With location information of each segment obtained in Step 1, the amplitude of residual noise can be estimated around each segment. For each segment, using the minimum value of $N_b$ points immediately before the segment and the minimum value of $N_a$ points immediately after the segment, the amplitude of the local residual noise can be estimated by averaging the two minimum values.  The residual noise removal can be obtained by subtracting the segment by its associated local residual noise amplitude. In our algorithm $N_b = N_a = 5$.

\subsection{Step 4: Classifying the Type of Each Segment}

At this point, there are only a few remaining segments in the filtered aggregated signal. And every segment can be classified into one of three types.
\begin{description}
  \item[Type 0:] The segment belongs to a dryer/oven waveform, or belongs to an EV waveform fully overlapping with a dryer/oven waveform which has almost the same duration as the EV waveform. For the former case, the segment can be simply removed since it is not an EV waveform. For the latter case, the segment should have  very high amplitude since a dryer/oven waveform has  high amplitude like an EV waveform (generally higher than 5 kW).

  \item[Type 1:] The segment belongs to an EV waveform,  or an AC lump, or an EV waveform overlapping with waveforms of non-AC appliances with relatively shorter durations, or an AC lump overlapping with waveforms of other appliances. One can calculate the approximate width and height of the segment, decide whether it is an EV waveform,  and then reconstruct the EV waveform.

  \item[Type 2:] The segment belongs to an EV waveform overlapping with an AC waveform, which is probably also overlapping with other appliances' waveforms. For example, the first two segments shown in Fig.\ref{fig:bumptrainfilter}(c) are respectively an EV waveform overlapping with an AC spike train and an EV waveform overlapping with an AC lump and a dryer waveform.
\end{description}

To classify a given segment $S(t)$ after Step 3, the following cumulative counting function is calculated:
\begin{eqnarray}
f( c ) = \langle  S(t) > c \rangle
\end{eqnarray}
where $c$ is an amplitude threshold from 0 to $\max(S(t))$, and the operator $\langle  S(t) > c \rangle$ counts the number of sampling points in $S(t)$ with an amplitude greater than $c$. For example, if $c=0$, then $f(c)$ is the total number of all nonzero samples in the segment. If $c=\max(S(t))$, then $f(c)=0$.

When calculating the gradient of the cumulative function $f( c )$, one can find that there are two \emph{prominent} peaks for Type 2 segments. This is because both an AC waveform and an EV waveform can be approximated as square waves, and a square wave can result in sharp drop in $f(c)$ when $c$ is equal to the height of the square wave. Similarly, there is one \emph{prominent} peak in the gradient of $f(c)$ for Type 1 segments and no \emph{prominent} peak for Type 0 segments. Thus, the number of prominent peaks in the gradient of the function suggests which type an observing segment belongs to.

\begin{figure}[t]
\begin{minipage}[b]{.48\linewidth}
  \centering
  \centerline{\epsfig{figure=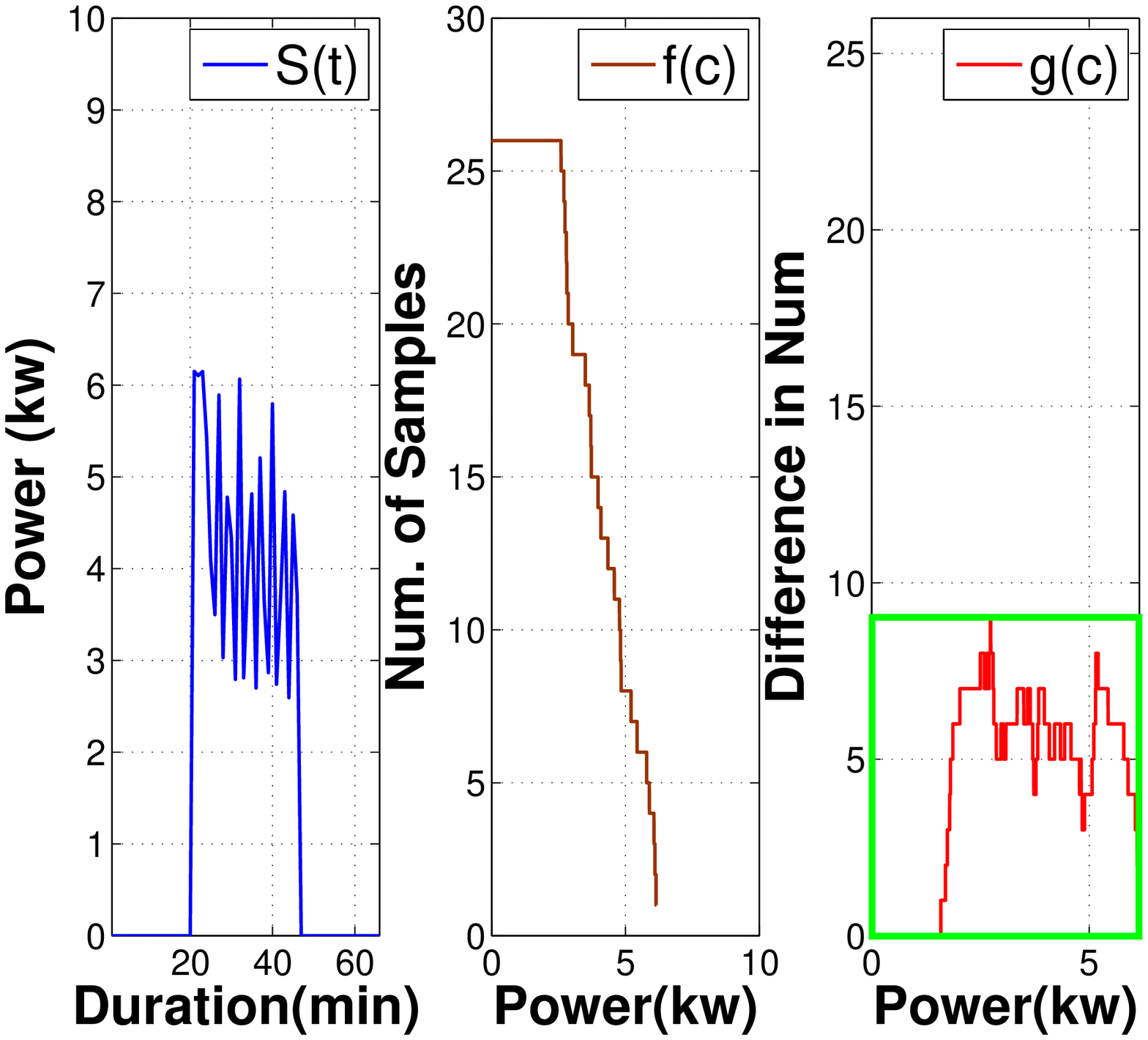,width=4.5cm,height=4cm}}
  \centerline{\footnotesize{(a)}}
\end{minipage}
\hfill
\begin{minipage}[b]{0.48\linewidth}
  \centering
  \centerline{\epsfig{figure=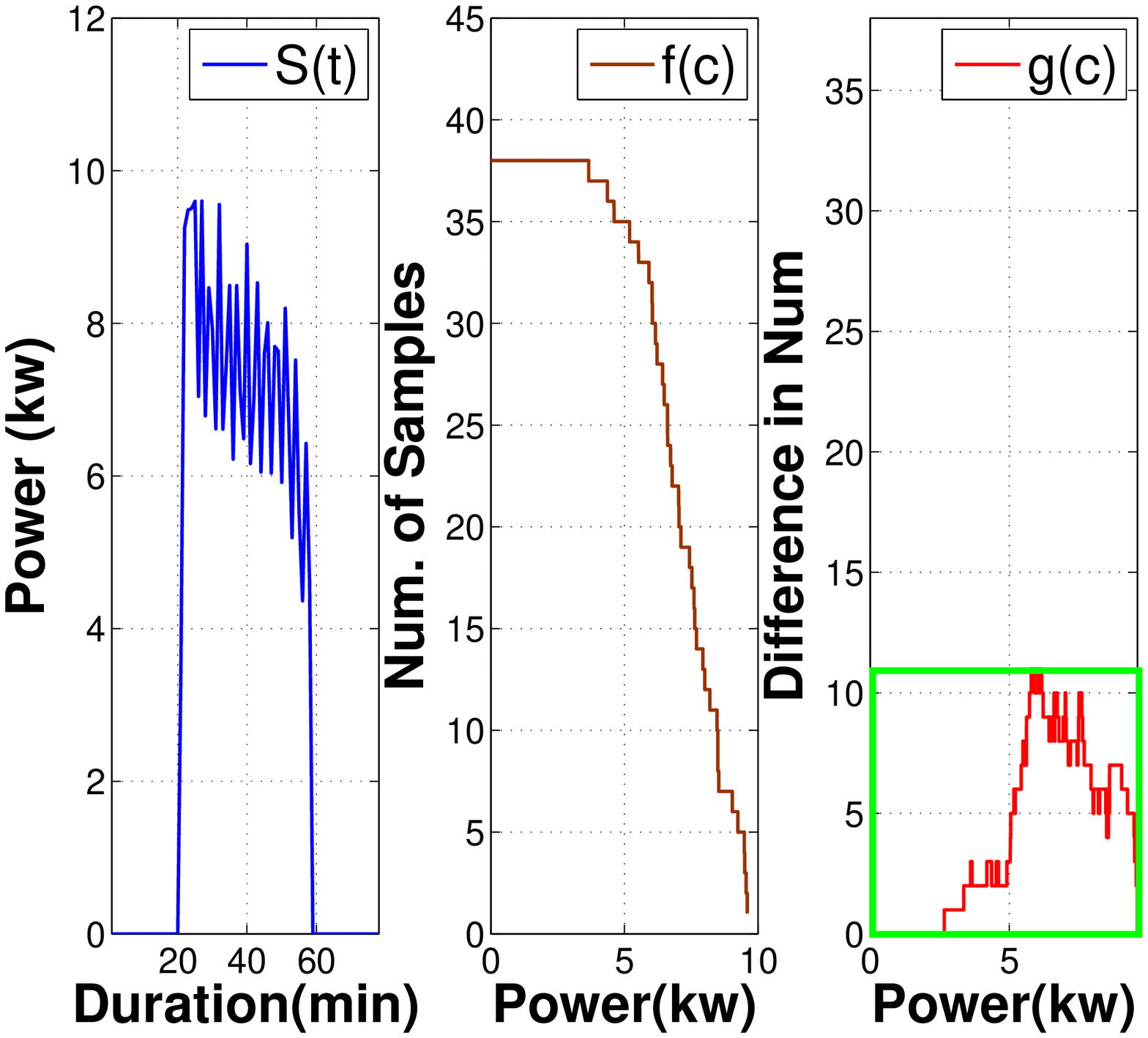,width=4.5cm,height=4cm}}
  \centerline{\footnotesize{(b)}}
\end{minipage}
\caption{\footnotesize{Two typical segments of Type 0. (a) shows a dryer wave. (b) shows an EV wave completely overlapped by a dryer wave which has the same duration as the EV wave. For each segment, the gradient of its cumulative function (shown in the middle plot in (a) and (b)) does not show prominent peaks (see the right plot in (a) and (b)).}}
\label{fig:type0}
\end{figure}

\begin{figure}[!t]
\begin{minipage}[b]{.48\linewidth}
  \centering
  \centerline{\epsfig{figure=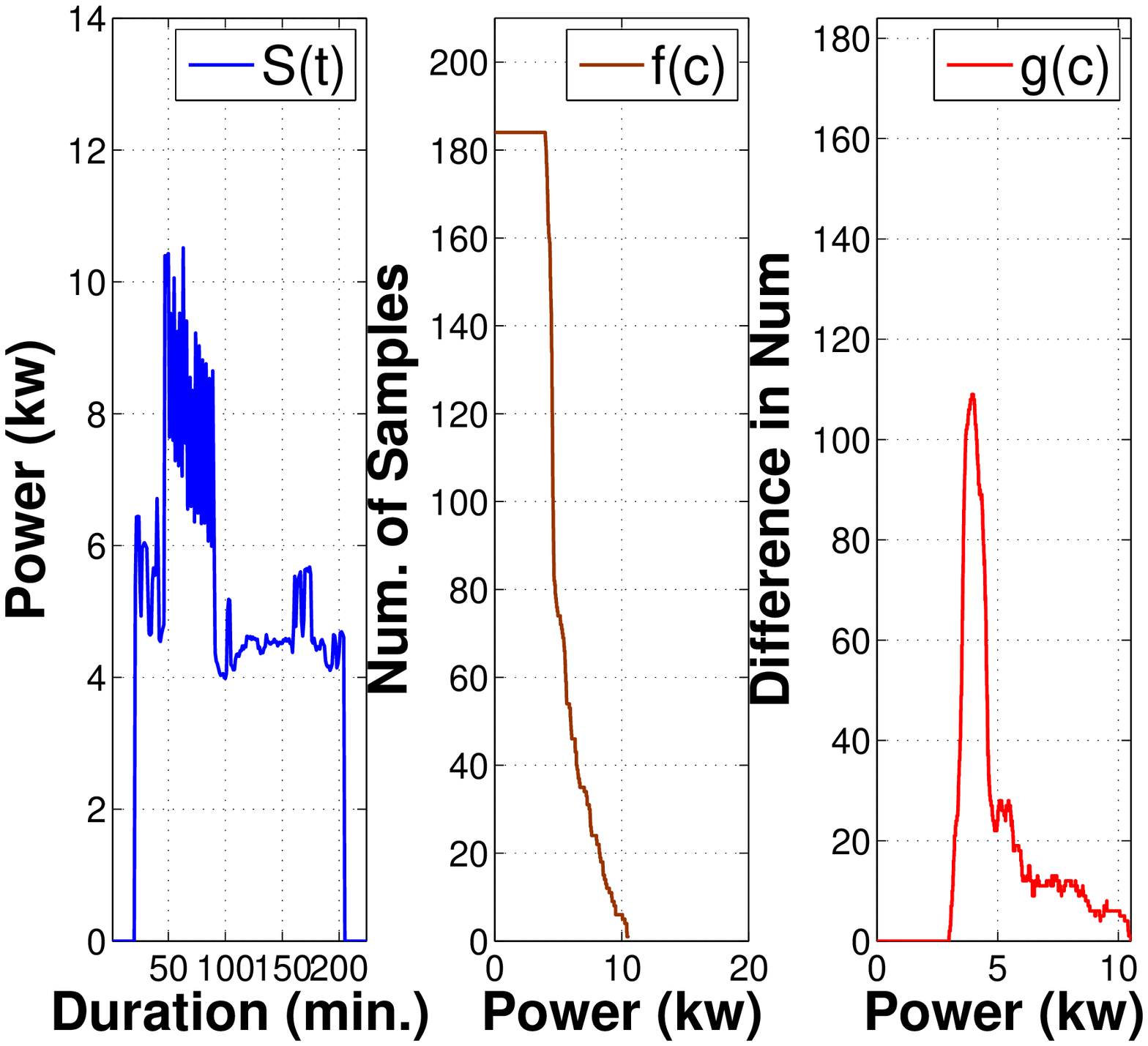,width=4.5cm,height=4cm}}
  \centerline{\footnotesize{(a)}}
\end{minipage}
\hfill
\begin{minipage}[b]{0.48\linewidth}
  \centering
  \centerline{\epsfig{figure=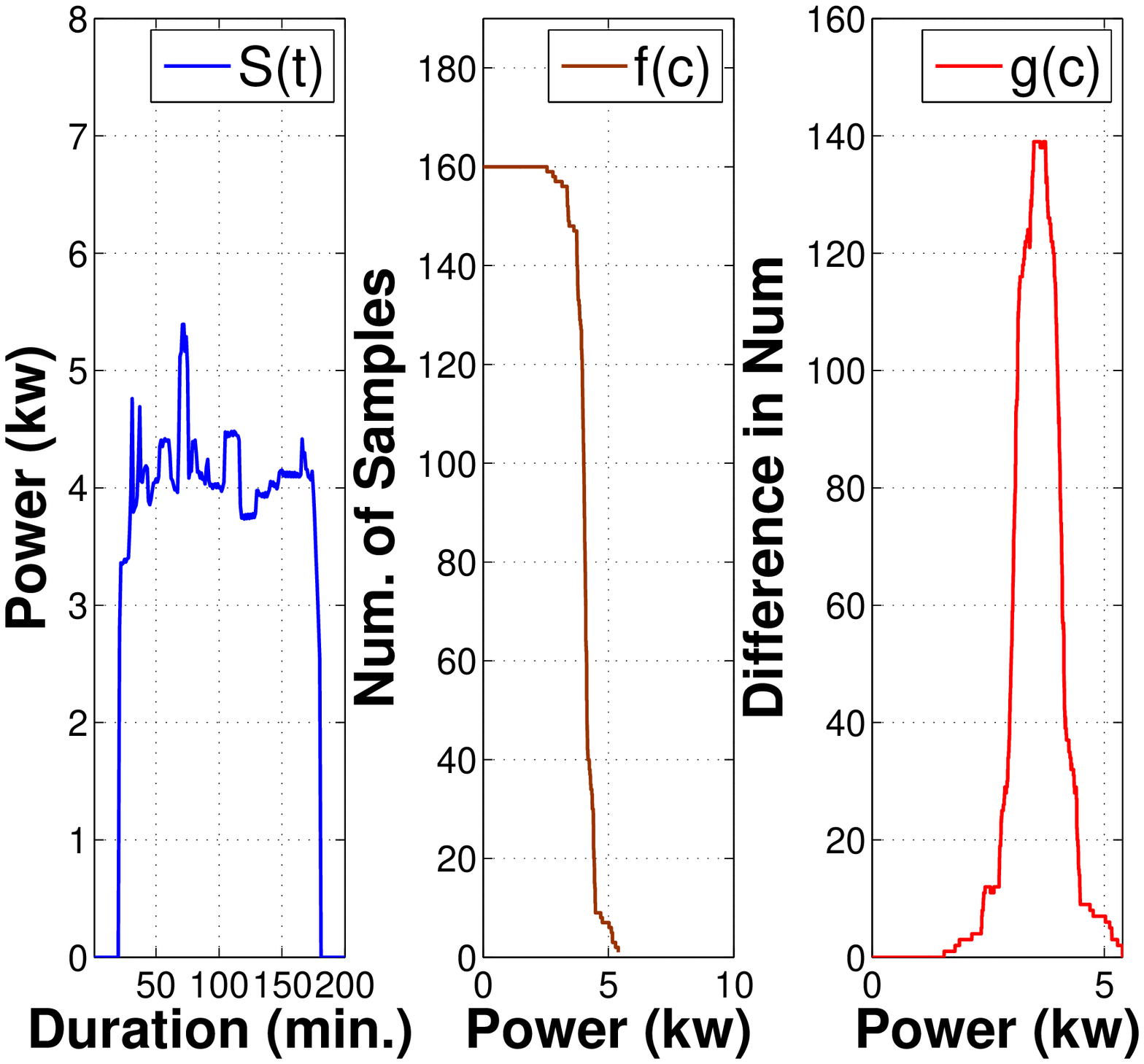,width=4.5cm,height=4cm}}
  \centerline{\footnotesize{(b)}}
\end{minipage}
\caption{\footnotesize{Two typical segments of Type 1. (a) shows an EV wave overlapped by a dryer wave with short duration. (b) shows an EV wave contaminated by fluctuation of residual noise. For each segment, the gradient of its cumulative function shows one prominent peak (shown in the right plot in (a) and (b)).}}
\label{fig:type1}
\end{figure}

\begin{figure}[!t]
\begin{minipage}[b]{.48\linewidth}
  \centering
  \centerline{\epsfig{figure=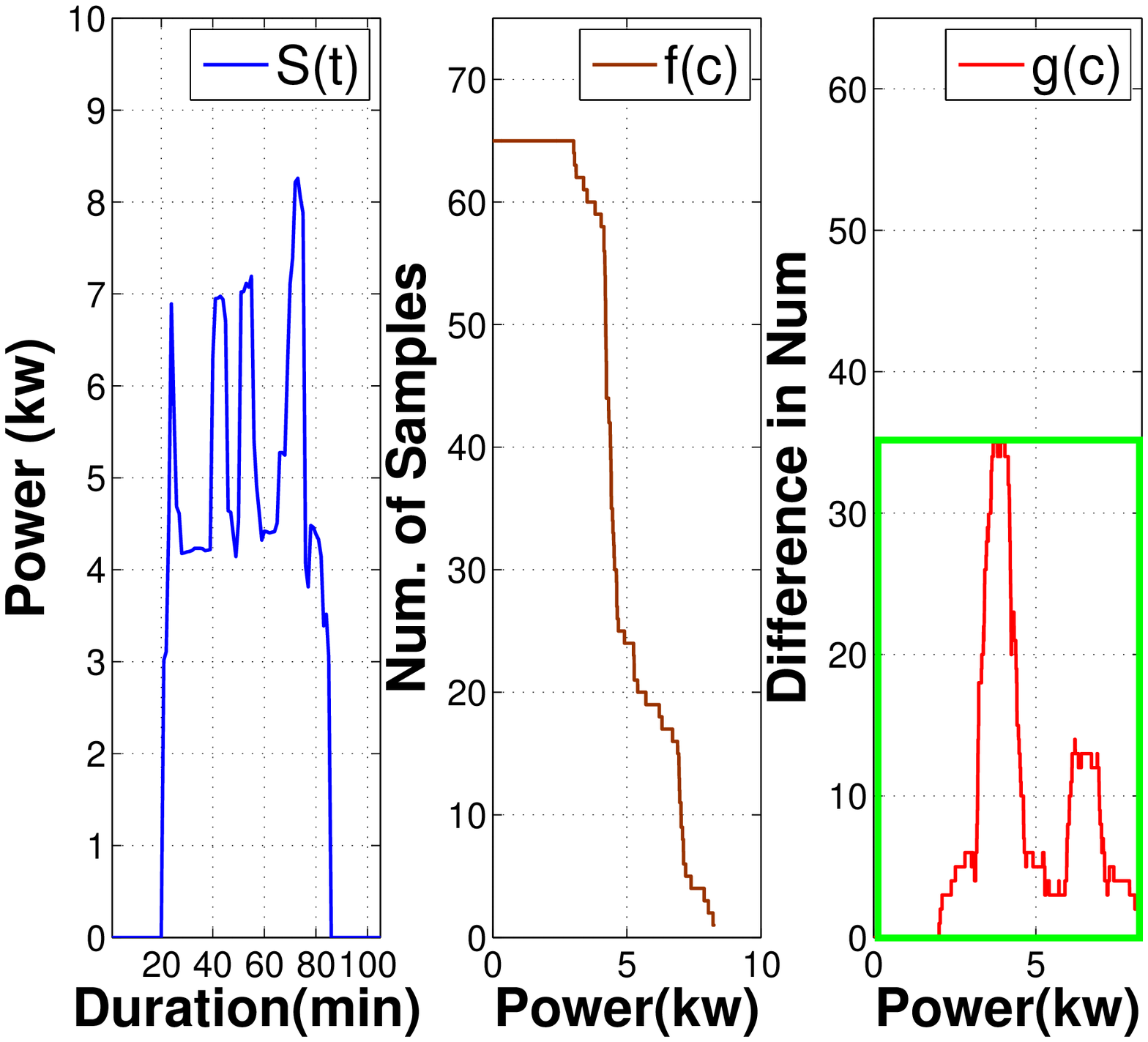,width=4.5cm,height=4cm}}
  \centerline{\footnotesize{(a)}}
\end{minipage}
\hfill
\begin{minipage}[b]{0.48\linewidth}
  \centering
  \centerline{\epsfig{figure=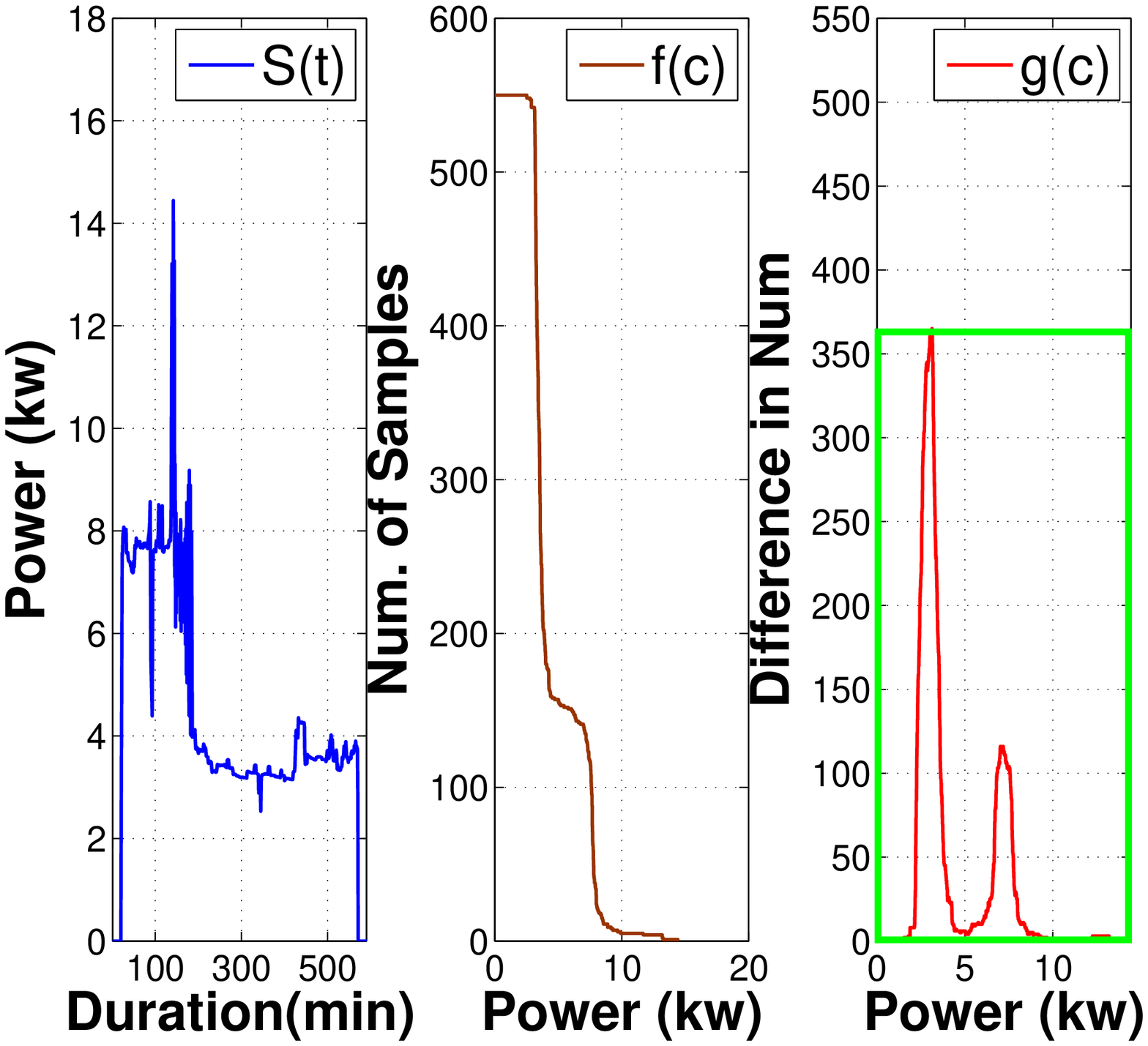,width=4.5cm,height=4cm}}
  \centerline{\footnotesize{(b)}}
\end{minipage}
\caption{\footnotesize{Two typical segments of Type 2. (a) shows an EV wave overlapped by an AC spike train, where the EV wave is the bottom part of the segment. (b) shows an EV wave overlapped by an AC lump and a dryer wave, where the EV wave is in the top part of the segment. The two segments are the first two segments in Fig.\ref{fig:bumptrainfilter}(c). For each segment, the gradient of its cumulative function shows two prominent peaks (shown in the right plot in (a) and (b)).}}
\label{fig:type2}
\end{figure}

To find \emph{prominent} peaks, we search peaks with mutual distance larger than 2 kW and peak height larger than $0.2\max(g)$ where $g$ is the gradient of $f$. The Matlab command \emph{findpeaks} can finish this task easily. If there is one peak, the segment is classified as Type 1 (Fig.\ref{fig:type1}). If there are at least two peaks, further calculate the area under the normalized gradient function $g_n \triangleq g/\max(g)$. If the area is larger than 35\% of the square area with the same width and height as $g_n$ (e.g. the green square area in Fig.\ref{fig:type0} and Fig.\ref{fig:type2}), then the segment is classified as Type 0 (Fig.\ref{fig:type0}); otherwise, it is classified as Type 2 (Fig.\ref{fig:type2}).  Examples of segments of Type 0, Type 1, and Type 2 are given in Fig.\ref{fig:type0}, Fig.\ref{fig:type1}, and Fig.\ref{fig:type2}, respectively.

\begin{figure}[!t]
\begin{minipage}[b]{.48\linewidth}
  \centering
  \centerline{\epsfig{figure=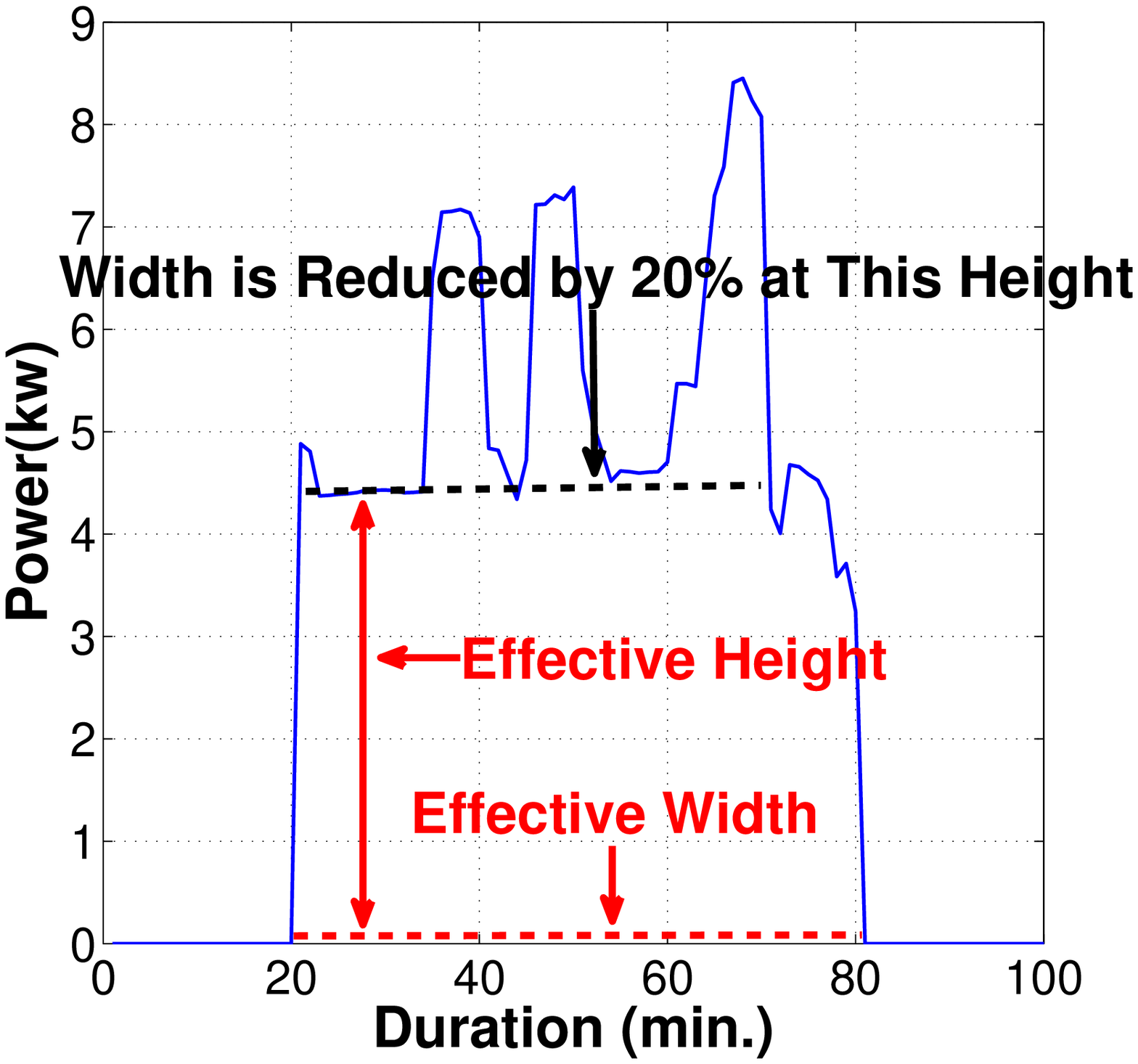,width=4.5cm,height=3.3cm}}
  \centerline{\footnotesize{(a) A segment}}
\end{minipage}
\hfill
\begin{minipage}[b]{0.48\linewidth}
  \centering
  \centerline{\epsfig{figure=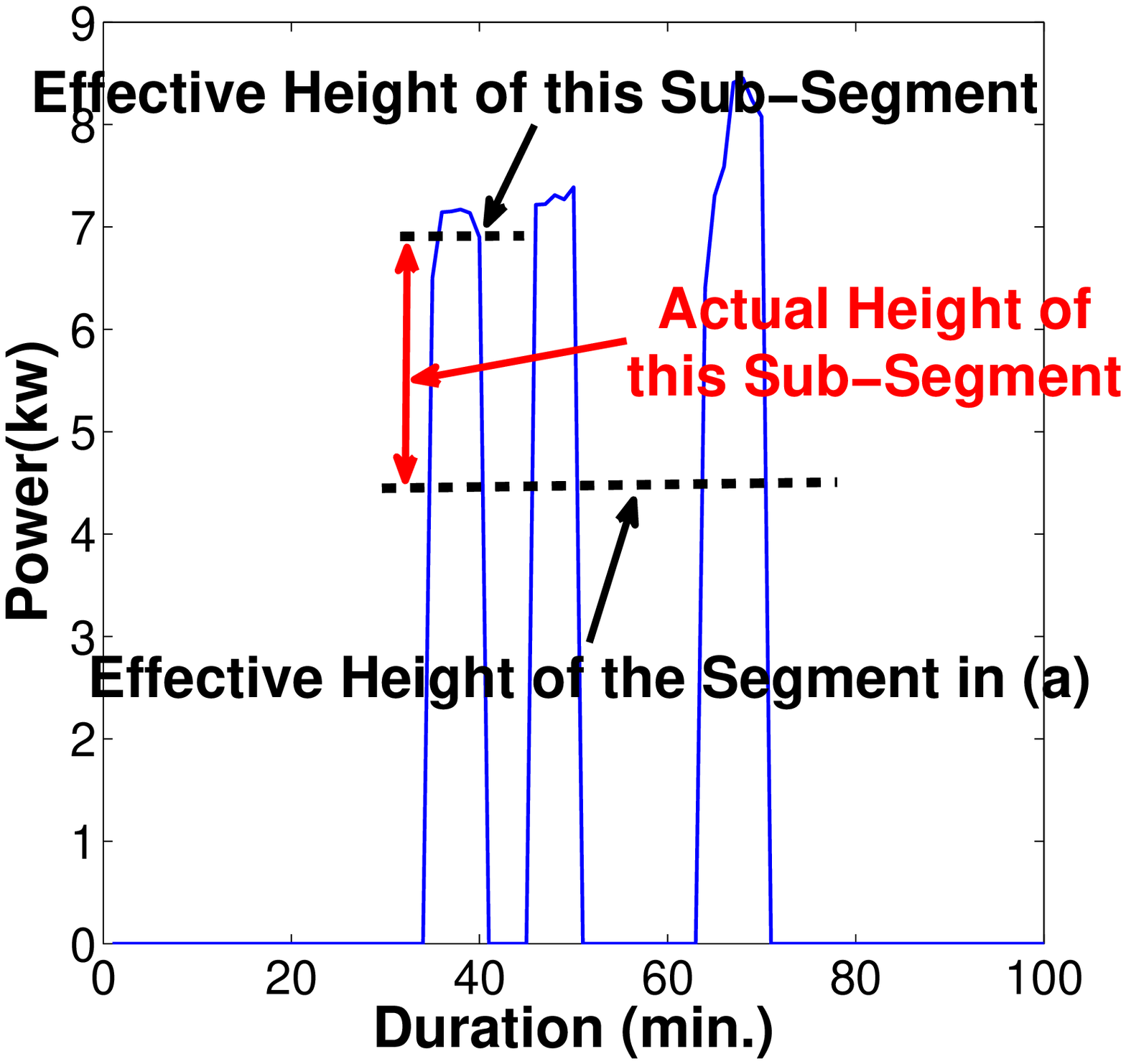,width=4.5cm,height=3.3cm}}
  \centerline{\footnotesize{(b) Sub-segments of the segment in (a)}}
\end{minipage}
\caption{\footnotesize{(a) Calculation of the effective width and height from a segment. (b) Calculation of the actual height of sub-segments of the segment in (a). The sub-segments are obtained by thresholding the segment with $T_{\mathrm{high}}=6$(kw).}}
\label{fig:EV_width}
\end{figure}

\subsection{Step 5: Energy Disaggregation}

Let us first introduce definitions of the effective width and the effective height of a segment. The \emph{effective width} is defined as the width of a segment at bottom. The \emph{effective height} is defined as the height at which the segment's width becomes only 80\% of the bottom width. Fig.\ref{fig:EV_width}(a) illustrates the calculation of the effective height and width.

If a segment belongs to \textbf{Type 0}, then we first determine its effective height. If its effective height is smaller than 5.5 kW, this segment is classified as a wave of dryer/oven (e.g. Fig.\ref{fig:type0}(a)). If larger than 5.5 kW,  the segment is classified as a fully overlapping waveform of an EV and a dryer/oven (e.g. Fig.\ref{fig:type0}(b)). For the latter case, it is impossible to accurately estimate EV waveform's height. However, considering the fact that an EV waveform has  constant and very stable amplitude from day to day, the EV waveform height can be assigned with a height estimate at another time of the same day or another day. Thus an EV square wave is reconstructed using the height and the calculated effective width.

If a segment belongs to \textbf{Type 1}, the effective height and width can be simply calculated, and then its square waveform can be reconstructed accordingly. However, if the width is very large, e.g. larger than 250 (minutes), the segment will be removed since an EV waveform generally exhibits a constant amplitude for no more than 2-3 hours \cite{PecanReportEV}. More likely, these long waveforms could be AC lumps (see Fig.\ref{fig:EVest}(b)) or other appliances' waveforms. Besides, if a candidate waveform has an effective height lower than 3 kw or is surrounded by a number of AC spikes, then it is treated as an AC lump as well.

If a segment belongs to \textbf{Type 2}, this segment can be considered to include both an EV waveform and an AC waveform (see the definition of Type 2 Segments). Thus it needs to be determined whether an EV waveform occupies the top part or the bottom part of the segment. For the illustration, see Fig.\ref{fig:type2}. In order to identify and separate two overlapped waveforms, first an additional threshold $T_{\mathrm{high}}$  will be used to obtain the sub-segment information of the top part. For the proposed algorithm, $T_{\mathrm{high}}=T_{\mathrm{low}}+2.5$(kW) is set. Similar to Step 1, using this threshold a number of sub-segments in the top part can be obtained as shown in Fig.\ref{fig:EV_width}(b).

Next, the effective width of the segment is calculated. If the width is larger than 250 (minutes), then the bottom part is more likely to be an AC lump due to the EV duration characteristic mentioned before. Thus, EV waveforms belong to the top part. Subsequently, the effective width and the \emph{actual height} \footnote{The \emph{actual height} of a sub-segment is calculated as the effective height of the sub-segment subtracted by the effective height of the associated segment. See Fig.\ref{fig:EV_width}(b) for illustration.} of each sub-segment (with duration longer than 20 minutes) are calculated to reconstruct an EV square waveform.

If the width is less than 250 (minutes), then the sub-segments  are analyzed. The proposed spike-train filter is used to remove the sub-segments. As a result, the following two cases are considered. (1) If the filter can remove all sub-segments, then the top part is an AC spike train, while the bottom part is an EV waveform \footnote{The bottom part cannot be an AC lump since an AC lump and an AC spike train cannot be overlapped.}. We can calculate the effective height and width of the bottom part to reconstruct the EV waveform. (2) If the spike-train filter cannot remove all sub-segments, then each remained sub-segment needs to be analyzed one by one. The \emph{actual height} of each remained sub-segment and the effective height of the segment need to be calculated. Whichever (sub-segment's actual height or the segment's effective height) is closer to an estimated EV height at another time of non-overlapping observation, it will be identified as an EV waveform \footnote{Of course, this cannot ensure the correct location of the EV waveform, considering errors in estimating the effective height and the actual height. However, in most cases, an EV waveform with satisfactory accuracy can be reconstructed since an EV waveform height is generally ranging from 3 kW to 4 kW while the height of an AC lump is generally smaller than 3 kW.}.

%
%

%

\subsection{Remarks}

Admittedly, the proposed algorithm uses a number of default values such as the amplitude and width of EV charging load signals. However, it is worthy emphasizing that these default values are based on general knowledge of EV charging load characteristics, and do not rely on a specific type of EV. For example, although the amplitude of EV charging load signals is changing from house to house, the amplitude is always larger than 3 kW. The proposed algorithm utilizes the amplitude range information, but not any exact amplitude number.

In the next section the proposed algorithm will be applied to a number of houses with robust performance across different houses and different seasons. This indicates the default values used in the algorithm do not affect practical use.

\begin{table*}[thp]
\caption{Performance comparison of our proposed algorithm and the HMM algorithm in \cite{parson2012non}. The last row of the table gives the performance (mean $\pm$ standard variance) averaged over all months and all houses.}
\label{Table1}
\centering
\begin{tabular}{c|c|c|c|c| c|c|c}
\hline\hline
\textbf{House }            &  \textbf{Month Range}  &  $\mathrm{\textbf{Err1}}$ (new)    &   $\mathrm{\textbf{Err2}}$ (new)        & $\mathrm{\textbf{MSE}}$ (new)    & $\mathrm{\textbf{Err1}}$ \cite{parson2012non}   &$\mathrm{\textbf{Err2}}$ \cite{parson2012non}  &  $\mathrm{\textbf{MSE}}$ \cite{parson2012non} \\
\hline
370                    & 2012-10 to 2013-09     & 8.0\%       & 11.7 (kwh) & 0.31   &        135.5\%           &192.2 (kwh) &    1.51        \\ \hline
545                   & 2012-09 to 2013-09   &  5.6\%        & 10.8 (kwh) &  0.13     &    89.1\%          & 156.8 (kwh)  &    1.06             \\    \hline
1782                   &   2012-05 to 2013-09    &   7.0\%      & 13.9 (kwh) &  0.17      &     28.8\%          & 79.3 (kwh)  &   0.42           \\ \hline
1801                  &  2012-07 to 2013-08  & 11.7\%         & 24.5 (kwh) &   0.29    &    76.2\%        & 156.5 (kwh) &   0.96            \\ \hline
2335                   & 2012-06 to 2013-05     &  9.6\%     & 20.3 (kwh) & 0.30    &       26.0\%           &58.0 (kwh) &     0.47    \\  \hline
3036                  & 2012-08 to 2013-09    &  5.9\%         & 20.3 (kwh) &  0.12     &     3.9\%         &12.9 (kwh) &   0.17          \\ \hline
3367                  & 2012-11 to 2013-10   &   5.9\%        & 9.9 (kwh) &  0.16     &        47.6\%       &81.3 (kwh) &  0.63              \\  \hline
6139                  & 2012-10 to 2012-05    &  10.1\%       & 20.9 (kwh) &  0.05       &      2.5\%        &5.0 (kwh) &    0.09        \\ \hline
7863                   &  2012-09 to 2013-09    &  9.2\%     & 21.1 (kwh) &    0.08        &    101.2\%       &   236.0 (kwh) & 1.02            \\ \hline
8669                   &   2012-09 to 2013-08    &  3.1\%     & 8.6 (kwh) &   0.15      &    26.7\%           &78.4 (kwh) &  0.30   \\   \hline
9934                   &  2012-10 to 2013-10   &  7.0\%      & 12.3 (kwh) &  0.27       &       38.8\%       & 73.1 (kwh) &  0.46          \\  \hline\hline
\textbf{Total}    &  125 months            &  \textcolor{red}{\textbf{7.5\% $\pm$ 6.2\%}}     &  \textcolor{red}{\textbf{15.7 $\pm$ 13.3} (kwh)} &  \textcolor{red}{\textbf{0.19 $\pm$ 0.15 }}      &     \textcolor{blue}{55.6\% $\pm$ 86.9\%}       & \textcolor{blue}{107.4  $\pm$ 163.5 (kwh)}     &  \textcolor{blue}{ 0.68 $\pm$ 0.97}      \\
\hline\hline
\end{tabular}
\end{table*}

\section{Experimental Results}
\label{sec:experiments}

An experiment was carried out to test performance of our proposed algorithm \footnote{Matlab codes are available at \url{https://sites.google.com/site/researchbyzhang/nilm}.}. For comparison, the HMM algorithm proposed in \cite{parson2012non} was used.

The data came from the Pecan Street Database \cite{PecanStreetDatabase}, which collects raw power signals recorded from hundreds of residual houses in Austin, Texas. Eleven houses using EV were randomly chosen from the database. Each house data contain aggregated power signals of about one year. Each aggregated power signal is generally a combination of about twenty power signals of various appliances, such as EV, AC, furnace, dryer, oven, range, dishwasher, cloth-washer, refrigerator, microwave, bedroom-lighting, and bathroom-lighting. The ground-truth power signals of these appliances are also available in the database. Thus the database is very suitable to test algorithms' performance in practice.

The eleven houses are listed in Table \ref{Table1}. Note that some houses have wrong ground-truth of EV power signals or bad recordings of aggregated signals in some months. Thus we remove the data of these months \footnote{The removed data are: House 545 in June 2013, House 1782 in July to September of 2012 and June to July of 2013, House 1801 in June to July of 2013, House 3036 in September to October of 2012 and June to July of 2013, House 3367 in June of 2013, House 7863 in June of 2013, and House 8669 in June of 2013.}. The remained data have total 125 months. The sampling rate is 1/60 Hz.

Since the HMM algorithm requires training, for each house its ground-truth power signals of EV and AC of two weeks were used as the training set. Note that our proposed algorithm does not need this training period.

Three performance indexes were used. One is the averaged estimation error of monthly energy consumption, defined as
\begin{eqnarray}
\mathrm{Err1} =  \frac{1}{N}\sum_{i}\sum_{j} \frac{|E^{i,j}_{\mathrm{true}} - E^{i,j}_{\mathrm{est}}|}{E^{i,j}_{\mathrm{true}}} \times 100\%
\end{eqnarray}
where $E^{i,j}_{\mathrm{true}}$ is energy consumption of the ground-truth EV power signal in the $j$-th month of the $i$-th year, and $E^{i,j}_{\mathrm{est}}$ is energy consumption of the estimated EV power signal in the same month, and $N$ is the total month number in the calculation.

A related performance index is the averaged estimation error of monthly energy consumption in kwh, defined as
\begin{eqnarray}
\mathrm{Err2} =  \frac{1}{N}\sum_{i}\sum_{j} |E^{i,j}_{\mathrm{true}} - E^{i,j}_{\mathrm{est}}| \quad (\mathrm{kwh})
\end{eqnarray}

The third performance index is the averaged normalized mean square error (MSE) in estimating EV charging load signals, defined as
\begin{eqnarray}
\mathrm{MSE} = \frac{1}{N}\sum_{i}\sum_{j} \frac{(X^{i,j}_{\mathrm{true}} - X^{i,j}_{\mathrm{est}})^2}{(X^{i,j}_{\mathrm{true}})^2}
\end{eqnarray}
where $X^{i,j}_{\mathrm{true}}$ is the ground-truth EV charging load signal in the $j$-th month of the $i$-th year, and $X^{i,j}_{\mathrm{est}}$ is the estimated EV charging load signal in the same month.

The results are presented in Table \ref{Table1}, which shows that the proposed algorithm significantly outperforms the HMM algorithm. For the proposed algorithm, the averaged estimation error of monthly energy consumption is only 7.5\%. Or put in another way, the error is only 15.7 kwh/month in average. In this experiment, the averaged monthly energy consumption of EV charging load is 208.5 kwh/month, and the averaged monthly total energy consumption of a house is 1109.9 kwh/month. Therefore, the estimation error of the proposed algorithm is well acceptable.

Using the average US electricity price in 2013 \cite{price}, i.e. \$0.117/kwh, the difference between the estimated monthly energy consumption and the ground-truth is only \$1.83/month, which is a small error, since the monthly energy consumption of EV charging is \$24.39/month in average and the monthly total energy consumption of a house is \$129.86/month in average.

In contrast, for the HMM algorithm, the averaged estimation error of monthly energy consumption is 55.6\%, or \$12.56/month.

In fact, the poor performance of the HMM algorithm is mainly due to the estimation error in summer, when AC becomes the strongest interference. To clearly see this, Table \ref{Table2} shows the performance averaged over all houses and the four summer months, namely June to September. From the results, one can see the HMM algorithm does not provide any meaningful estimation for these four months; the averaged estimation error of monthly energy consumption of EV charging is 152.7\%, or \$34.11/month, and the averaged normalized MSE is 1.81. (A meaningful disaggregation result should have normalized MSE much smaller than 1.)

\begin{figure}[t]
\centering
\includegraphics[width=8cm,height=8cm]{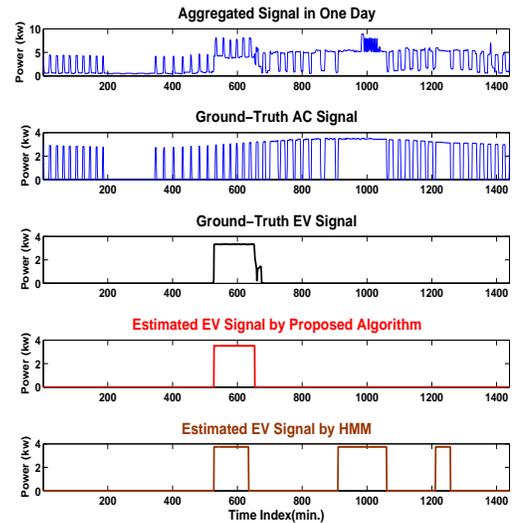}
\caption{One example showing the estimation performance of our proposed algorithm and the HMM algorithm in \cite{parson2012non}. From top to down, there are an aggregated power signal of one day, the ground-truth of AC power signal, the ground-truth of EV charging load signal, the estimated EV charging load signal by our proposed algorithm (in red color), and the estimated EV charging load signal by the HMM algorithm. It is seen that the HMM algorithm treats some AC lumps as parts of the EV charging load signal. In contrast, the proposed algorithm correctly disaggregates the EV signal.}
\label{fig:result2}
\end{figure}

Fig.\ref{fig:result2} shows an example of the estimated EV charging load signals by the two algorithms. One can see the HMM algorithm treats some AC lumps as parts of the EV signal, and thus makes large errors. In contrast, the proposed algorithm correctly identifies and disaggregates the EV charging load signals from the aggregated signals.

\begin{table}[t]
\caption{Performance (mean $\pm$ standard variance) of the proposed algorithm and the HMM algorithm in \cite{parson2012non} averaged over all houses and the summer months (June, July, August, and September). }
\label{Table2}
\centering
\begin{tabular}{c|c|c|c}
\hline\hline
            &     $\mathrm{\textbf{Err1}}$      &   $\mathrm{\textbf{Err2}}$        & $\mathrm{\textbf{MSE}}$     \\
\hline
New Algorithm              &   \textcolor{red}{\textbf{7.4\% $\pm$ 6.6 \%}}       & \textcolor{red}{\textbf{16.1 $\pm$ 15.7} (kwh) }        & \textcolor{red}{\textbf{0.28 $\pm$ 0.19} }          \\ \hline
HMM \cite{parson2012non}   &    \textcolor{blue}{152.7\% $\pm$ 114.4\% }      & \textcolor{blue}{ 291.5 $\pm$ 213.5 (kwh)}   &  \textcolor{blue}{1.81 $\pm$ 1.21 }                  \\
\hline\hline
\end{tabular}
\end{table}

%
%

\section{Conclusions}

In this paper, a new algorithm was proposed for non-intrusive energy disaggregation of electric vehicle charging given a real aggregated power signal. The new algorithm does not require training, demands a light computational load, and renders a high energy estimation accuracy.  These advantages were illustrated by experiments on the real world data with a low sampling rate (1/60 Hz) delivering superior performance even under the presence of air-conditioners.

\bibliographystyle{IEEEtran}

\bibliography{bib_NILM}

\end{document}